\documentclass[twocolumn,showpacs,preprintnumbers,amsmath,%
amssymb,prl,superscriptaddress]{revtex4}

\usepackage{graphicx}
\usepackage[pdfsubject={Overdoped cuprates' electronic structure},%
pdfauthor={D.C. Peets},pdftitle={X-ray Absorption Reveals Collapse %
of Single-Band Hubbard Physics in Overdoped Cuprates},%
pdfpagetransition=Glitter]{hyperref}

\def\Tc{\ensuremath{T_c}}
\def\TBCO{Tl$_2$Ba$_2$CuO$_{6+\delta}$}
\def\YBCO{YBa$_2$Cu$_3$O$_{x}$}
\def\LSCO{La$_{2-x}$Sr$_x$CuO$_{4\pm\delta}$}
\def\Eperpc{\ensuremath{E\perp c}}

\begin{document}

\title{Measurement   of   x-ray   absorption  spectra   of   overdoped
  high-temperature  cuprate  superconductors:  Inapplicability of  the
  single-band Hubbard model}


\author{D.C.~Peets}
\email{dpeets@scphys.kyoto-u.ac.jp}
\affiliation{Department of Physics, Graduate School of Science, Kyoto 
University, Kyoto, Japan 606-8502}

\author{D.G.~Hawthorn}
\affiliation{Department of Physics \& Astronomy, University of 
Waterloo, 200 University Ave.~W, Waterloo, ON, Canada N2L 3G1}

\author{K.M.~Shen}
\affiliation{Department of Physics, Cornell University, Ithaca, New 
York, USA 14853}

\author{Young-June Kim}
\author{D.S.~Ellis}
\author{H.~Zhang}
\affiliation{Department of Physics, University of Toronto, 60 
St.~George St., Toronto, ON, Canada M5S 1A7}

\author{Seiki Komiya} 
\affiliation{Central Research Institute of Electric Power Industry, 
2-6-1 Nagasaka, Yokosuka, Kanagawa, Japan 240-0196} 
\author{Yoichi Ando} 
\affiliation{Institute of Scientific and Industrial Research, Osaka 
University, 8-1 Mihogaoka, Ibaraki, Osaka, Japan 567-0047}

\author{G.A.~Sawatzky} 
\author{Ruixing Liang} 
\author{D.A.~Bonn} 
\author{W.N.~Hardy} 
\affiliation{Department of Physics \& Astronomy, University of 
British Columbia, 6224 Agricultural Rd., Vancouver, BC, Canada 
V6T 1Z1} 
\affiliation{Canadian Institute for Advanced Research, Canada}

\date{\today}

\begin{abstract}

X-ray   absorption   spectra   on   the   overdoped   high-temperature
superconductors     Tl$_2$Ba$_2$CuO$_{6+\delta}$     (Tl-2201)     and
La$_{2-x}$Sr$_x$CuO$_{4\pm\delta}$ (LSCO)  reveal a striking departure
in the electronic  structure from that of the  underdoped regime.  The
upper Hubbard band, identified with strong correlation effects, is not
observed  on the  oxygen {\slshape  K} edge,  while  the lowest-energy
prepeak gains  less intensity  than expected above  $p\sim0.21$.  This
suggests  a breakdown of  the Zhang-Rice  singlet approximation  and a
loss  of  correlation effects  or  a  significant  shift in  the  most
fundamental  parameters of the  system, rendering  single-band Hubbard
models  inapplicable.   Such  fundamental  changes  suggest  that  the
overdoped regime  may offer a  distinct route to understanding  in the
cuprates.

\end{abstract}


\maketitle

The greatest  unsolved problem in correlated electron  physics is that
of high-temperature  superconductivity in the cuprates.   As holes are
doped into  the cuprates' Mott-insulating parent  compound, a plethora
of  new electronic  phases  emerge \cite{Lee2006},  but  for only  the
undoped Mott insulator does  an agreed-upon microscopic picture exist.
The remaining phases are characterized by unusual electronic structure
--- the    Fermi    surface    may    consist    of    remnant    arcs
\cite{Norman1998,Shen2005,Kanigel2006} or pockets enclosing only a few
percent           of            the           Brillouin           zone
\cite{Doiron2007,leBoeuf2007,Bangura2008,Jaudet2008},
superconductivity  with unconventional  $d_{x^2-y^2}$-symmetry pairing
emerges      \cite{Muller1986,Scalapino1995,Tsuei2000},      and     a
nonsuperconducting  `pseudogap' state  exhibits a  gap  reminiscent of
superconducting phases.

Past optimal doping,  where the superconducting transition temperature
\Tc\  peaks, the  materials  begin to  behave  more like  conventional
metals.   This  `overdoped' side's  resistivity  approaches the  Fermi
liquid  $T^2$ power  law  \cite{Mackenzie1996,Nakamae2003}, while  the
Fermi      surfaces       of      overdoped      \TBCO\      (Tl-2201)
\cite{Hussey2003,Peets2007}   and  \LSCO\   (LSCO)  \cite{Yoshida2006}
closely   resemble    those   from   band    structure   calculations.
Understanding  this  evolution  may  prove key  to  understanding  the
cuprates' phase diagram.

An  undoped  cuprate's  CuO$_2$-plane  copper  atoms  have  electronic
configuration  [Ar]3$d^9$, with  one  hole in  each Cu  $3d_{x^2-y^2}$
orbital;  although band  structure calculations  in the  local density
approximation (LDA) predict metallic  behavior, it is instead a charge
transfer insulator  with a 2--3~eV gap ---  strong correlation effects
produce  upper  and  lower  Hubbard bands  corresponding  to  electron
addition and  removal states of the system  respectively, separated in
energy by $U$, with an oxygen  band between them.  The doping of holes
into  this correlated insulator  manifests as  a transfer  of spectral
weight from the Hubbard bands into states near the Fermi energy, often
labeled the `Zhang-Rice singlet band' \cite{ZhangRice1988}.  Both one-
and  three-band   models  calculate   this  transfer  to   be  roughly
proportional                         to                         doping
\cite{Eskes1991,Chen1991,Hybertsen1992,Meinders1993},   in   agreement
with  x-ray absorption measurements  at the  oxygen {\slshape  K} edge
(probing O  $2p$ unoccupied states) on underdoped  and optimally doped
cuprates            \cite{Chen1991,Chen1992,Pellegrin1993a,Nucker1995}.
Measurements   on  a   limited  number   of  overdoped   LSCO  samples
\cite{Chen1992,Pellegrin1993a},  however,  suggested  the  low  energy
spectral weight to plateau on further doping, and optical measurements
have also observed changes in overdoped LSCO \cite{Uchida1991}.  Since
preparation of overdoped LSCO  crystals without oxygen deficiencies or
inhomogeneity  is difficult,  it  may be  tempting  to attribute  this
inconsistency  to poor  sample quality.   If verified,  however, these
results could  herald a significant  changes in the  cuprates' physics
and a  fundamental breakdown of  the approaches most commonly  used to
model them.

Although the doped holes occupy oxygen orbitals, they can reduce their
energy by hybridizing with a  Cu spin to form a $d_{x^2-y^2}$-symmetry
state --- the Zhang-Rice singlet \cite{ZhangRice1988,Eskes1988}.  Each
doped hole locally  compensates a local copper spin  and thus mimics a
doped  hole  in  a  single  band  Hubbard  model  with  $3d_{x^2-y^2}$
symmetry,   as   opposed  to   a   three-band   model  comprising   Cu
$3d_{x^2-y^2}$  and  O  $2p_x$  and  $2p_y$  states.   This  effective
single-band description  underpins the widely used $t-J$  model of the
cuprates.

In  this Letter,  we report  new x-ray  absorption  spectroscopy (XAS)
measurements at  the oxygen  {\slshape K} edge  in overdoped  LSCO and
Tl-2201, against the backdrop of underdoped \YBCO\ (YBCO).  We confirm
the  weakening  of  low  energy  spectral weight  transfer  at  higher
dopings; that  this is  observed in both  families indicates it  to be
generic to the overdoped cuprates.   We argue that these results point
to  a fundamental  change  in the  electronic  structure past  optimal
doping.   This has  bearing  on  effective models  of  the low  energy
physics in  the cuprates, suggesting the  inapplicability in overdoped
materials  of a  Zhang-Rice description  or any  model based  upon it,
placing  these materials  firmly in  uncharted  theoretical territory.
These  changes  may  help  explain  the comparative  normalcy  of  the
overdoped regime,  and suggest  that it may  offer a new  and distinct
starting point for understanding in the cuprates.


Single  crystals  of  \TBCO\  were  grown  by  a  self-flux  technique
\cite{Peets2007}   and  annealed   under  controlled   oxygen  partial
pressures   to  obtain   the  desired   doping  levels   and  maximize
homogeneity.   The  crystals  were  etched  using  dilute  bromine  in
anhydrous ethanol  to ensure  high-quality surfaces, then  masked with
gold to  eliminate any  spurious contributions.  Crystals  were stored
and  transported in  H$_2$O- and  CO$_2$-free environments  to prevent
surface  damage.   Tl-2201's  doping  was estimated  using  Presland's
phenomenological formula \cite{Presland1991},
\begin{equation}\label{eq:presland}
1 - \frac{T_c}{T_c^{max}} = 82.6(p-0.16)^2
\end{equation}
with  $T_c^{max}=94$~K,  which  has  proven  consistent  with  dopings
determined from  the Luttinger theorem  and measurements of  the Fermi
surface   area  in   heavily  overdoped   Tl-2201  crystals   by  AMRO
\cite{Hussey2003} and ARPES \cite{Peets2007}.

Samples  of La$_2$CuO$_4$ \cite{Ono2007}  and LSCO  were grown  by the
traveling  solvent floating  zone technique,  cut to  expose  the $ac$
plane, polished  with 0.05~$\mu$m alumina grit and  etched with dilute
Br  prior  to  measurement.   Samples  of  YBCO  were  oxygen-annealed
self-flux-grown single  crystals \cite{Liang1998} with polished/etched
$ac$  or as-grown  $ab$ plane  surfaces; doping  was determined  as in
Ref.~\cite{Liang2006}.

Oxygen  {\slshape K} edge  X-ray absorption  spectra were  measured at
beamline 8.0.1  of the Advanced  Light Source \cite{Jia1995},  with an
energy  resolution  of 0.40~eV,  via  total  fluorescence yield  (TFY,
significantly less  surface-sensitive than total  electron yield), and
normalized by incident intensity.  For Tl-2201, spectra were collected
at room temperature  on overdoped crystals with $T_c$s  of 9.5~K, 60~K
and 69~K for \Eperpc; a  \Tc~=~60~K crystal measured with a 40$^\circ$
angle between $E$ and $a$, such that all features were visible, showed
no  change between  91~K and  room  temperature.  LSCO  and YBCO  were
measured at normal incidence with $E \parallel a$.

To account for  variations in signal strength (e.g.\  less of the beam
strikes smaller samples), the  spectra were further normalized against
each other  using energy  ranges below and  well above the  main edge.
Changes in  the beamline's energy calibration  introduced small energy
shifts, precluding an overall calibration; instead, the spectra's main
absorption    edges    were   shifted    to    match   earlier    work
\cite{Pellegrin1993b,Chen1992,Nucker1995}.   The key  results reported
in this Letter are  insensitive to the transitions' absolute energies,
which are a  function of the initial core level's  energy and can vary
by crystallographic site and orbital symmetry.


\begin{figure}[htb]
\includegraphics[width=\columnwidth]{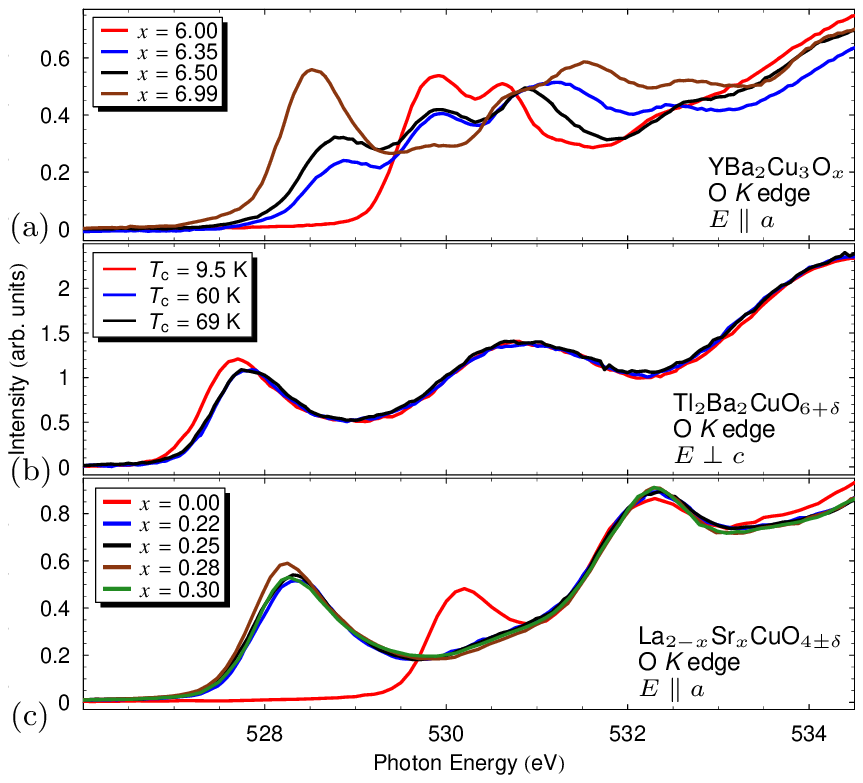}
\caption{\label{fig:OK_edge}(Color   online)   Doping  dependence,   O
  {\slshape K} edge, $E\parallel a$: (a) YBCO shows typical underdoped
  cuprate  behavior  ---  the  lowest-energy prepeak,  absent  in  the
  undoped ($x=6.00$) cuprate, grows rapidly and shifts to lower energy
  as  doping increases; the  next prepeak,  identified with  the upper
  Hubbard band,  is suppressed, but  remains clearly visible  to light
  overdoping ($x=6.99$).   (b) In Tl-2201,  the upper Hubbard  band is
  not observed,  while the  lowest-energy prepeak exhibits  only minor
  doping dependence.  (c) Overdoped LSCO's doping dependence resembles
  Tl-2201's; undoped $x=0$ provides a reference point.}
\end{figure}

YBCO, Tl-2201 and  LSCO in-plane oxygen {\slshape K}  edge spectra are
shown in  Fig.\ \ref{fig:OK_edge}.  From undoped  ($x=6.00$) to slight
overdoping ($x=6.99$), YBCO's lowest energy pre-edge peak (prepeak) at
$\sim$528.3  eV decreases  in energy  and strengthens  while  the next
prepeak  ($\sim529.8$  eV, identified  with  the  upper Hubbard  band)
weakens,  consistent   with  theoretical  expectations   and  previous
measurements        \cite{Nucker1995,Pellegrin1993a}.        Tl-2201's
lowest-energy prepeak  is only visible for \Eperpc,  identifying it as
an extension  of that in YBCO;  it continues to  increase in intensity
and  decrease in energy  with doping,  although its  intensity changes
more  gradually.  Similar  behavior is  seen in  LSCO.   The remaining
prepeaks  in Tl-2201  are  attributed  to the  BaO  and primarily  the
Tl$_2$O$_2$ layers, by their polarization-dependence and by comparison
to  LDA band structure  calculations (not  shown).  The  upper Hubbard
band, ubiquitous in underdoped  cuprates and clearly observed in YBCO,
is  absent or  weak  and nearly  doping-independent  in the  overdoped
materials' oxygen edges.

\begin{figure}[htb]
\includegraphics[width=\columnwidth]{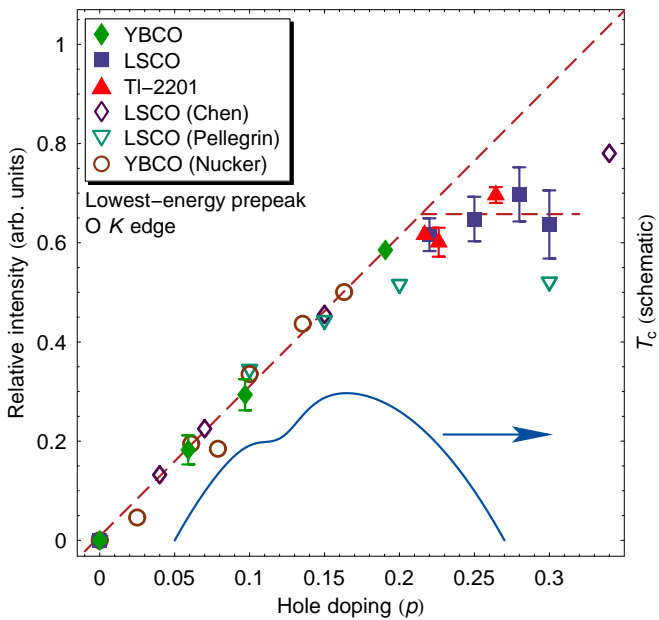}
\caption{\label{fig:weight}(Color   online)   Doping   dependence   of
  lowest-energy  O {\slshape  K} edge  prepeak ($E\parallel  a$); data
  from N\"ucker \cite{Nucker1995},  Chen \cite{Chen1992} and Pellegrin
  \cite{Pellegrin1993a}  have been  included  and the  superconducting
  dome is provided for reference.  The weight closely tracks $p$ up to
  and  past optimal doping  ($p=0.16$), but  deviates from  this trend
  around  $p=0.21$, exhibiting  much weaker  doping dependence  in the
  strongly overdoped regime.}
\end{figure}

The   lowest-energy   prepeak's   doping   dependence  is   shown   in
Fig.\   \ref{fig:weight}  for   Tl-2201,  YBCO   and   LSCO;  previous
measurements  are  included.    Integration  windows  were  526.75  --
529.25~eV  (YBCO), 526.7  -- 529.4~eV  (LSCO), and  525.6  -- 529.4~eV
(Tl-2201); weights  were normalized by  assuming LSCO and  YBCO follow
the same  line in the underdoped  regime and by  equating the $x=0.22$
LSCO and Tl-2201 $p=0.217$ ($\Tc=69$~K) weights.  That YBCO, a bilayer
material, should  exactly track single-layer LSCO is  not obvious, but
this is immaterial  to the overdoped trends discussed  below.  A small
offset of $\sim$0.02 due to non-zero weight in the undoped spectra was
subtracted  from the  LSCO and  YBCO weights,  which assumes  a linear
doping  dependence  of  the  Hubbard band.   The  prepeak's  intensity
increases roughly linearly with doping  over much of the doping range,
consistent  with  one  or  three  band  Hubbard  descriptions  of  XAS
\cite{Eskes1991,Meinders1993,Chen1991,Hybertsen1992}.          However,
overdoped Tl-2201  and LSCO deviate  from this trend  around $p=0.21$,
exhibiting weaker  doping dependence  and possibly even  decreasing at
the highest LSCO doping.   This clear departure, apparently generic to
overdoped  cuprates,  conflicts  with  the basic  description  of  the
cuprates  in a doped  one or  three band  Hubbard model;  indeed, this
behavior more closely resembles that  of a simple band metal, with one
new Fermi level hole state per  doped hole rather than the two derived
from  spectral  weight  transfer.   Together with  the  missing  upper
Hubbard band, these  results indicate a clear change  in the nature of
the electronic structure.

These changes  likely mark a  fundamental breakdown of  the Zhang-Rice
singlet model at  high doping levels and with  it the applicability of
single band Hubbard or $t-J$ models.  At $p=0.25$, the probability for
nearest  neighbor Zhang-Rice  singlets reaches  unity.  These  share a
common oxygen atom, making them  non-orthogonal and costing the O 2$p$
on-site repulsion $U_{pp}$.  With Zhang-Rice singlets thus unstable at
high dopings  and inapplicable  as a basis  state, one  key unanswered
question is what takes their place.  This limitation of the theory may
be obvious, at  least in retrospect, but very  little theoretical work
has  considered it,  and what  could  extend or  supplant the  current
single-band Hubbard approaches remains an open question.

In a  correlated picture, at high  doping levels the high  O 2$p$ hole
density   would  increase  the   influence  of   $U_{pp}$  ($\sim5$~eV
\cite{vanderMarel1988} and similar to  the O 2$p$ bandwidth), reducing
Cu--O covalency  in the ground state  \cite{Macridin2005}.  This would
{\slshape intensify} strong correlation  effects for Cu 3$d$ holes and
further  stabilize local  Cu spins.   The upper  Hubbard band  is only
visible on the oxygen edge due to strong Cu 3$d$ -- O 2$p$ mixing, and
would quickly  vanish if covalency fell.  Eskes  showed the low-energy
dynamical spectral weight transfer to depend strongly on the covalency
$t_{pd}/\Delta_{pd}$ (the ratio of  the intersite hopping parameter to
the charge transfer energy) \cite{Eskes1991,CTGapNote}, so a reduction
in covalency  would strongly  attenuate features in  the O  2$p$ doped
hole region.  Leaving the region where  Zhang-Rice singlet-like states
are stable, the system may enter the  region of free O 2$p$ holes in a
Cu  ($S=1/2$) lattice background  \cite{Eskes1988}.  The  greater role
for $U_{pp}$  would increase the charge transfer  energy and introduce
correlation effects  to the oxygen  bands.  Oxygen holes  would retain
their mainly $p_x$ and $p_y$ character (these states remain atop the O
$2p$  bands),   but  Zhang-Rice  singlets  would   no  longer  enforce
$d_{x^2-y^2}$ symmetry.   The loss  of covalency would  greatly reduce
superexchange,  and with  little change  expected in  $U_{dd}$, copper
atoms would behave more like  isolated spins.  Data suggestive of such
behavior have been reported in overdoped LSCO \cite{Wakimoto2005}.


The  absence   of  spectral   features  associated  with   the  strong
correlation  effects ubiquitous  in  the underdoped  regime signals  a
fundamental  change  in  the  cuprates'  electronic  structure  around
$p=0.21$, although  the nature  of this change  is unclear  and futher
work will  be required  to determine its  origin, full  properties and
exact location.   These changes suggest  the breakdown of  single band
Hubbard  approaches  as  holes  cease  to be  dilute.   The  overdoped
cuprates  may  behave  as  simple,  ordinary  band  metals,  with  the
underdoped regime's Fermi arcs  and other curious electronic structure
features  being connected  to the  success of  the  Zhang-Rice singlet
model.   The correlations  could  instead be  masked  from the  oxygen
{\slshape K}  edge by  reduced covalency, with  an increased  role for
$U_{pp}$  and  significant   doping-dependent  shifts  in  fundamental
parameters of the system such  as the charge-transfer energy and $p-d$
hopping terms (different parameters are already required for electron-
and  hole-doped cuprates  \cite{Eskes1991}).   The latter  explanation
suggests Fermi  liquid-like O 2$p$ holes  weakly coupled to  a Cu 3$d$
spin  bath, reminiscent  of the  heavy Fermion  superconductors.  Both
explanations are inconsistent  with the Zhang-Rice singlet-like states
underpinning single band Hubbard models; fundamentally new theoretical
approaches   will  be   required  to   successfully  model   the  full
superconducting  dome.  Ultimately,  these  results may  tie into  the
trend   toward  more  conventional   Fermi  liquid-like   behavior  as
\Tc\ decreases in the overdoped  cuprates, and may open a fruitful new
route to understanding in these otherwise enigmatic materials.

\begin{acknowledgments}

The  authors wish  to  thank  J.D.\ Denlinger  and  I.S.\ Elfimov  for
assistance.  This  work was  supported by NSERC,  the CRC  program and
BCSI.  The Advanced Light Source  is supported by the Director, Office
of Science, Office  of Basic Energy Services, of  the US Department of
Energy  under Contract No.\  DE-AC02-05CH11231.  Y.A.~is  supported by
KAKENHI 20030004 and 19674002.

\end{acknowledgments}

\bibliographystyle{h-physrev}
\bibliography{dpeets_tlxas}

\begin{thebibliography}{10}
\bibitem{CTGapNote}
Note that the charge transfer energy $\Delta_{pd}$ between the band centers is
  distinct from the charge transfer gap seen in transport measurements.
\end{thebibliography}

\end{document}